\documentclass{PoS}

\title{Understanding Cygnus X-3 Through Multi-Wavelength Studies}

\ShortTitle{Understanding Cygnus X-3 Through Multi-Wavelength Studies}

\author{\speaker{Michael McCollough}%
         \thanks{This work is supported by NASA contract NAS8-03060.}\\
        Smithsonian Astrophysical Observatory, U.S.A.\\
        E-mail: \email{mmccollough@head.cfa.harvard.edu}}

\author{Karri Koljonen\\
        Aalto University Mets\"ahovi Radio Observatory, Finland\\
        E-mail: \email{karri@kurp.hut.fi}}

\author{Diana Hannikainen\\
        Aalto University Mets\"ahovi Radio Observatory/Finnish Centre for Astronomy with ESO, Finland\\
        E-mail: \email{diana@kurp.hut.fi}}

\author{Guy Pooley \\
        Astrophysics Group,Cavendish Laboratory, U.K.\\
        E-mail: \email{guy@mrao.cam.ac.uk}}

\author{Sergei A. Trushkin\\
        Special Astrophysical Observatory RAS, Russia\\
        E-mail: \email{sergei.trushkin@gmail.com}}

\author{Danny Steeghs\\
        University of Warwick, U.K.\\
        E-mail: \email{D.T.H.Steeghs@warwick.ac.uk}}

\author{Marco Tavani \\
        INAF-IASF, Italy\\
        E-mail: \email{marco.tavani@iasf-roma.inaf.it}}

\author{Robert Droulans \\
        CESR/CNRS - Universit\'e de Toulouse, France\\
        E-mail: \email{Robert.Droulans@cesr.fr}}

\abstract{Cygnus X-3 is a unique microquasar which shows X-ray state changes, 
strong radio emission, and relativistic jets.  
It is also an unusual X-ray binary with the mass-donating companion being a 
high mass star Wolf-Rayet but the orbital modulation (as inferred from X-ray 
emission) is only 4.8 hours, a value more common in low-mass systems.
It has recently been shown by AGILE and Fermi that
Cygnus X-3, is a transient gamma-ray source (>100 MeV).  

To understand the environment, nature, and behavior of Cygnus X-3
multi-wavelength observations are necessary.  
In this proceedings we present the results achieved so far from
multi-wavelength campaigns.}  

\FullConference{8th INTEGRAL Workshop ``The Restless Gamma-ray Universe''\\
		 September 27-30 2010\\
		 Dublin Castle, Dublin, Ireland}

\begin{document}

\section{Introduction}

Cygnus X-3 is a unique microquasar composed of a compact object and a 
Wolf-Rayet star \cite{vk1}.  
It has a strong 4.8 hour orbital modulation and is 
known to produce relativistic jets \cite{pgk,ma}.  
It also exhibits relatively strong radio emission most of the time 
($\rm \sim 100~mJy$) \cite{we}. {\it AGILE} \cite{tm} and {\it Fermi} \cite{fl} 
have recently shown that Cygnus X-3 is a transient gamma-ray source 
($\rm >~100~MeV$).  Flaring gamma-ray emission appears to be connected with 
major radio flares and their associated X-ray states that precede the major 
radio flares \cite{tm, mkh}.

Studies \cite{we,mm} in the radio and hard X-ray (HXR) have found four states: 
quiescent, minor flaring, quenched and major flaring. Table 1 shows how these 
states are correlated with the simultaneous X-ray observations.  Throughout all
states of activity the HXR and soft X-ray (SXR) show an anti-correlation or spectral
pivoting around 10 keV \cite{mm2}.

Additional studies have shown that the X-ray emission can be broken down into
additional states.  A study of the X-ray and radio behavior 
arrive at six states (quiescent, minor flaring, suppressed, post-flare, major 
flaring, and quenched) \cite{szm}.  A recent study which incorporated X-ray 
spectral hardness (shape of the spectrum) in addition to the X-ray and radio 
arrive at three 
major states, Quiescent, Flaring, and  Hypersoft, with several substates 
\cite{khm}.  The Hypersoft is a new 
state which has important ties to the quenched state, gamma-ray emission, and 
jet production.

\section{Results}

The multi-wavelength campaigns (2006/2010) initiated to study the environment, 
nature and behavior of Cygnus X-3 included data from space-based 
({\it INTEGRAL}, {\it AGILE}, {\it Chandra}, {\it Swift}, {\it RXTE}, and 
{\it Fermi}), ground-based radio (Ryle, RATAN-600, Mets\"ahovia, VLBA, VLA, 
and WSRT), and Infrared (PAIRITEL, Calar Alto) observatories.  Some of the
initial results of these campaigns are presented below.

\subsection{Multi-Wavelength Look at a Major Radio Flare}

{\it INTEGRAL} Target of Opportunity
Observations (TOOs) allowed a detailed look at Cygnus X-3's X-ray spectrum 
near the peak of major radio flares.  These included observations within 0.25 
days, 2.5 days, and 3.5 days of the flare peak in the radio.  In addition to 
{\it INTEGRAL} TOOs, observations were made with {\it RXTE}, {\it Swift}, and
ground-based radio and infrared observations (see \cite{m7iw,k7iw}).  This
allowed us to assemble Cygnus X-3 SEDs during these major radio flares. 

Initial fits with physical models such as Eqpair \cite{cp} and BELM  \cite{bmm} 
show: 
{\tt (a) Radio Spectrum:} At the peak 
of the 
radio flare the radio spectrum has become optically thin;
{\tt (b) Absorption:} In all of the observations there is strong absorption
from the infrared to the SXR;
{\tt (c) Model Components:} To model the data it was necessary to have a disk
component, a compact corona, and a diluted jet.  Most of the radio emission
arises from outside of the core region;
and {\tt (d) Optical Depth:}  As the flare evolves the optical depth of the core
of the system decreases.

Some additional major findings from these campaigns were:

{\tt Quenched State Trigger:} For all of the major radio flares examined it has
been found that Cygnus X-3 must pass through the quenched state 
\cite{szm,khm}.  Specifically it appears that Cygnus X-3 passing through the 
Hypersoft state may be crucial for the formation of major radio flares 
\cite{khm}. In order to determine if Cygnus X-3 is in a quenched state an
examination of both the SXR and HXR is necessary in addition to the
radio.  Quenching in the radio may not always be observed due to the 
radio emission from decaying earlier flares (post-flare state \cite{szm}).  

{\tt Correlation with the HXR:}  When Cygnus X-3 is in a flaring state
the HXR and radio are correlated with each other.  The HXR appears to lag the 
radio in the 2006 major flare between three to five days (see Fig. 1).  During 
the flaring state the HXR values are in general below those seen during 
quiescence.  

{\tt X-Ray Spectral Evolution:} The X-ray spectrum of Cygnus X-3
consistently evolves from a soft spectrum (Hypersoft) at the beginning of the
flare to become progressively harder as the flare peaks and decays (see Fig. 1).  
This is due to the presence of a HXR tail that initially increases in intensity 
and then becomes fainter and harder as the flare decays.  

{\tt Low Frequency QPOs:}  A recent discovery is that low frequency
QPOs (LFQPOs), first reported by \cite{vdk}, are common and associated 
with the major radio flares \cite{kk}.  The LFQPOs are found to follow a major 
radio flare and are observed only in the orbital phase range of $ \rm 0.2-0.7$ 
\cite{kk}.

\begin{figure}
\includegraphics[angle= 90.0,width=0.5\textwidth]{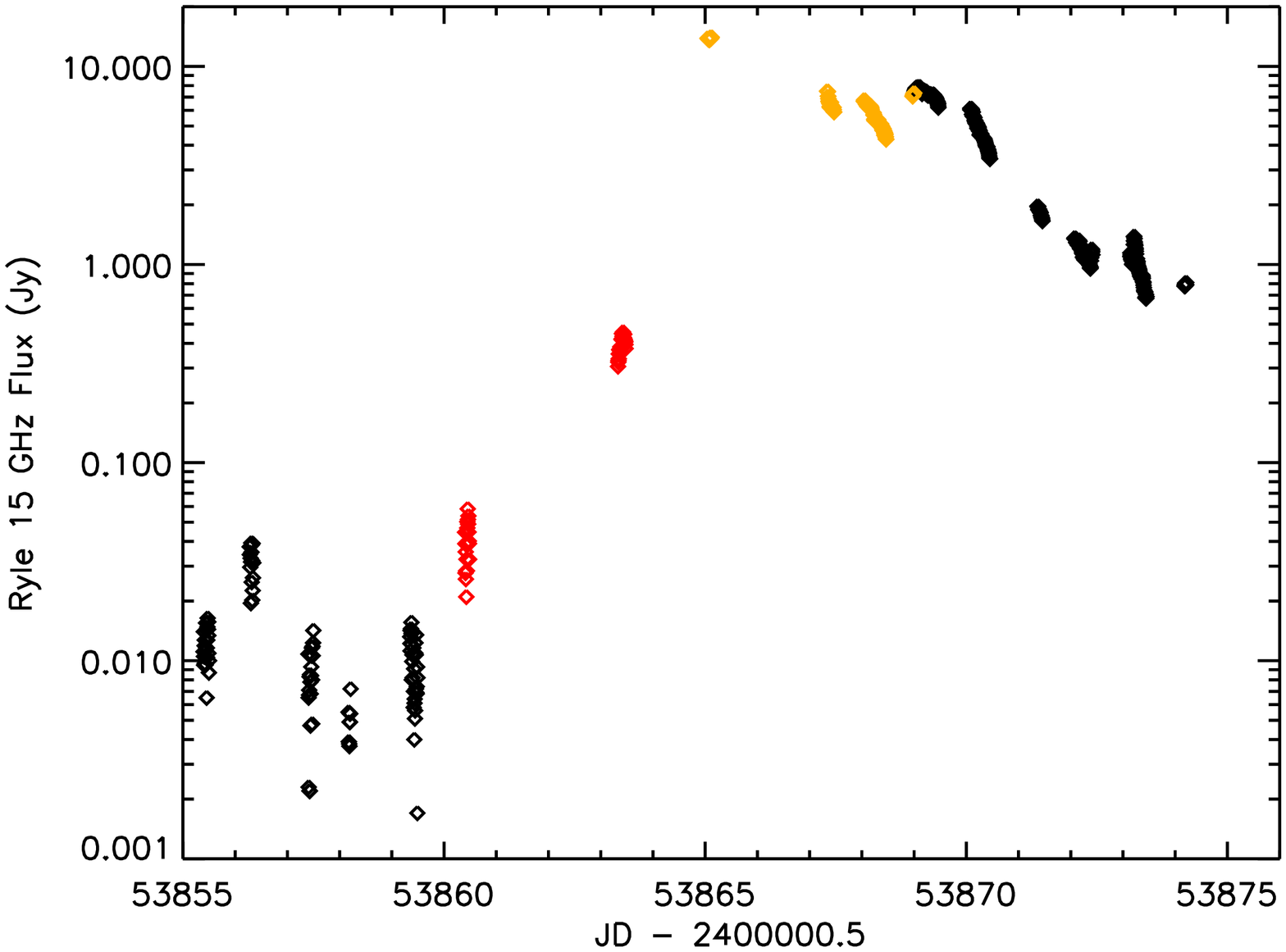}
\includegraphics[angle= 90.0,width=0.5\textwidth]{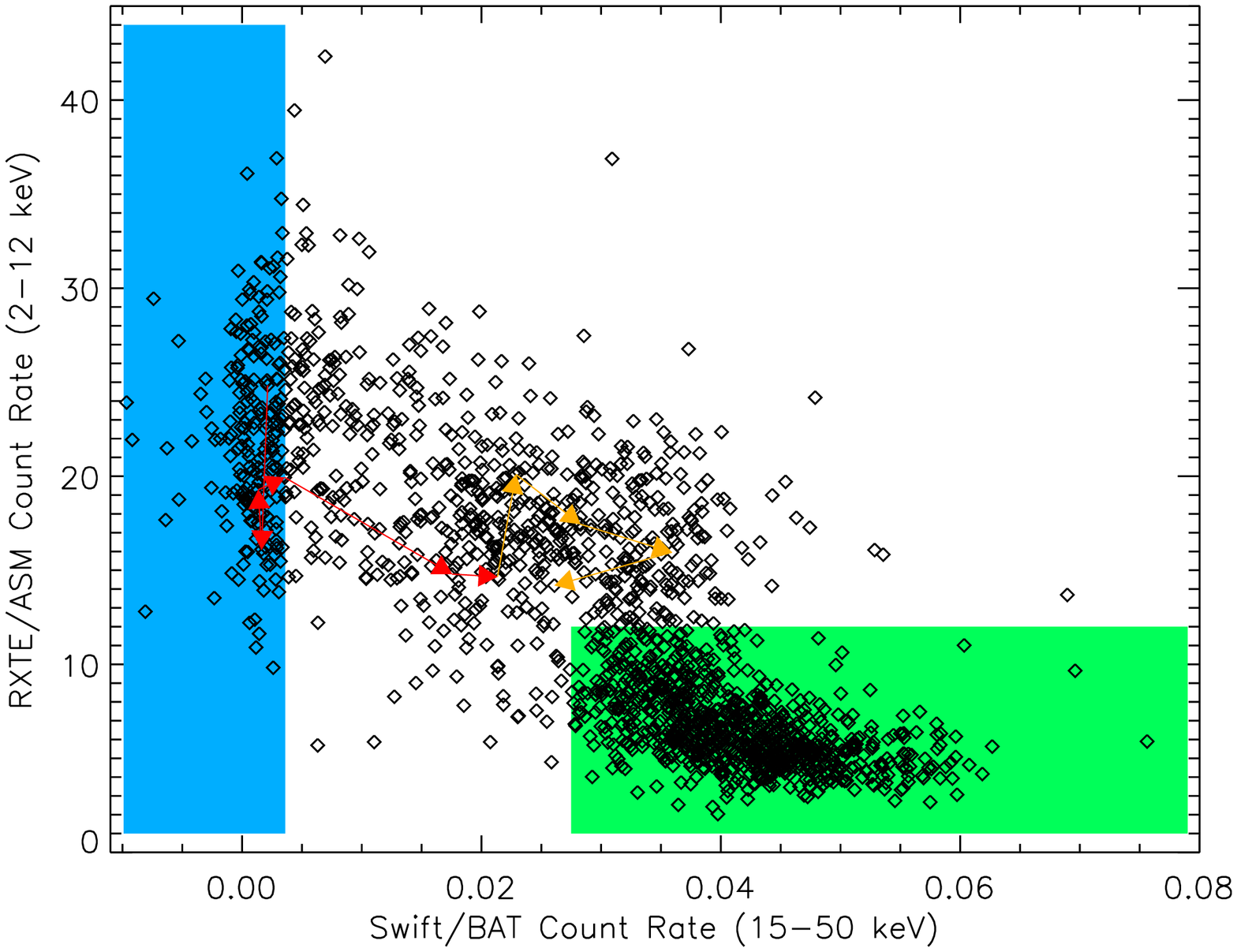}
\caption{{\it left:}  Radio light curve for the first major radio flare 
observed by 
{\it INTEGRAL}.  Note the color coding for before the peak (red) and after
(orange).
{\it right:} A HXR vs. SXR plot for Cygnus X-3.  The
path followed during the major radio flare are shown with red before the peak
and orange after.  The blue represents the quenched state and the green the
quiescent state.}
\label{fig1}
\end{figure}

\begin{table}
\begin{center}
\begin{tabular}{|c|c|c|c|c|c|}
\hline
State & Radio Level & SXR & HXR & Radio - SXR & Radio - HXR \\
\hline
 quiescent & $\rm \sim 60-100~mJy$ & low & high & + & - \\
\hline
 minor flaring &$\rm < 1~Jy$  & $\uparrow$ & $\downarrow$ & * & * \\
\hline
 quenched & $\rm \sim 1-15~mJy$ & high & low & - & + \\
\hline
 major flaring & $\rm \sim 1-20~Jy$ & $\downarrow$ & $\uparrow$ & - & + \\
\hline
\end{tabular}
\caption{For each of Cygnus X-3's states \cite{we,mm} is given the radio level,
SXR and HXR behavior (high, low, rising $\uparrow$ or falling $\downarrow$) and 
correlations between the radio and the SXR and HXR.  For the correlations + is 
a correlation and - is an anti-correlation.  The * represents where no 
correlation has been found.}
\label{tab1}
\end{center}
\end{table}

\subsection{Gamma-Ray Emission}

Cygnus X-3 has been detected by various space missions out to energies of 
$\rm \sim 200-300~keV$.  Throughout the 1970s and 1980s there were reported
detections of Cygnus X-3 in the MeV to TeV energy ranges.  But none of these
detections were confirmed or found reproducible.  Recent observations by 
{\it AGILE} and {\it Fermi} have shown that
the Cygnus region is a very complicated area at gamma-ray energies.
Cygnus X-3 has now been shown by both {\it AGILE} \cite{tm} and {\it Fermi}
\cite{fl} to be a gamma-ray source ($\rm > 100~MeV$).  

{\it AGILE} detected gamma-ray emission during periods of HXR quenching and 
very high SXR activity (typical Cygnus X-3 quenched state).
The gamma-ray spectrum between 100 MeV and 3 GeV is well described by a
power-law spectrum with a photon index of $1.8 \pm 0.2$.  The gamma-ray emission
appear to be correlated with the recently identified Hypersoft state \cite{khm}.  

{\it Fermi} found a $\rm 29~\sigma$ source consistent with Cygnus X-3.  
{\it Fermi} also found that using only data taken during periods of flaring 
activity, the gamma-ray emission exhibited the 4.8 hour orbital modulation 
associated with Cygnus X-3. Although it was found to be shifted by $0.3-0.4$ in 
phase relative what is observed at other wavelengths.

In May 2010 {\it AGILE} and {\it Fermi} both detected renewed gamma-ray emission
from Cygnus X-3 \cite{agile10}.  In both episodes of activity
the gamma-ray emission was associated with very brief periods of HXR quenching. 
In one case the HXR quenching lasted only $\rm \sim 0.5~day$ and in the other
for a 1 day.  In the last period of activity a 1 Jy radio flare (15 GHz) was
observed during the period of gamma-ray emission (see Fig. 2).

Various models have been proposed to explain the gamma-ray emission in terms of
leptonic \cite{dch} or hadronic processes \cite{rtkm}.

\begin{figure}
\begin{center}
\includegraphics[angle= 90.0,width=0.49\textwidth]{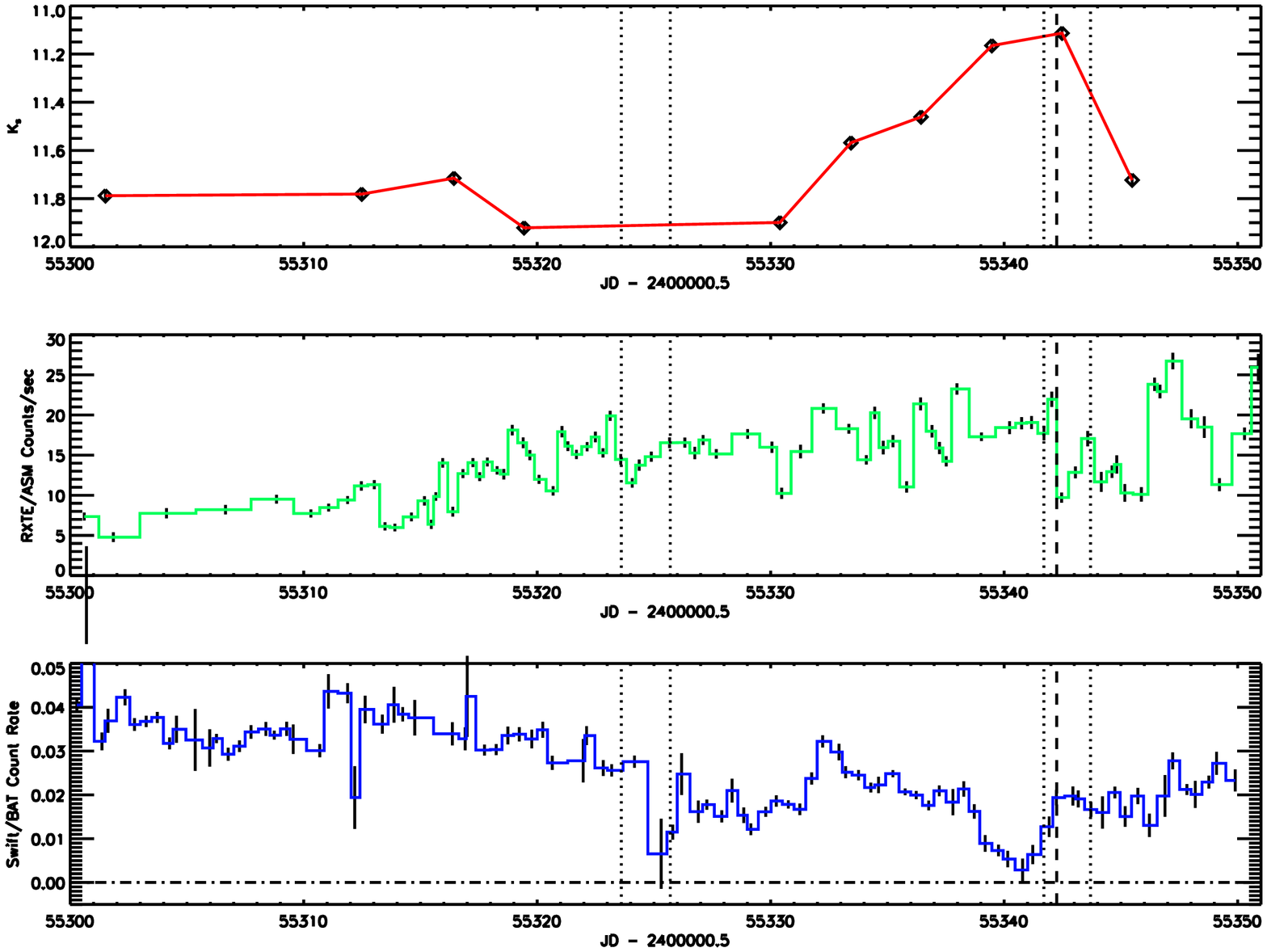}
\includegraphics[angle= 90.0,width=0.49\textwidth]{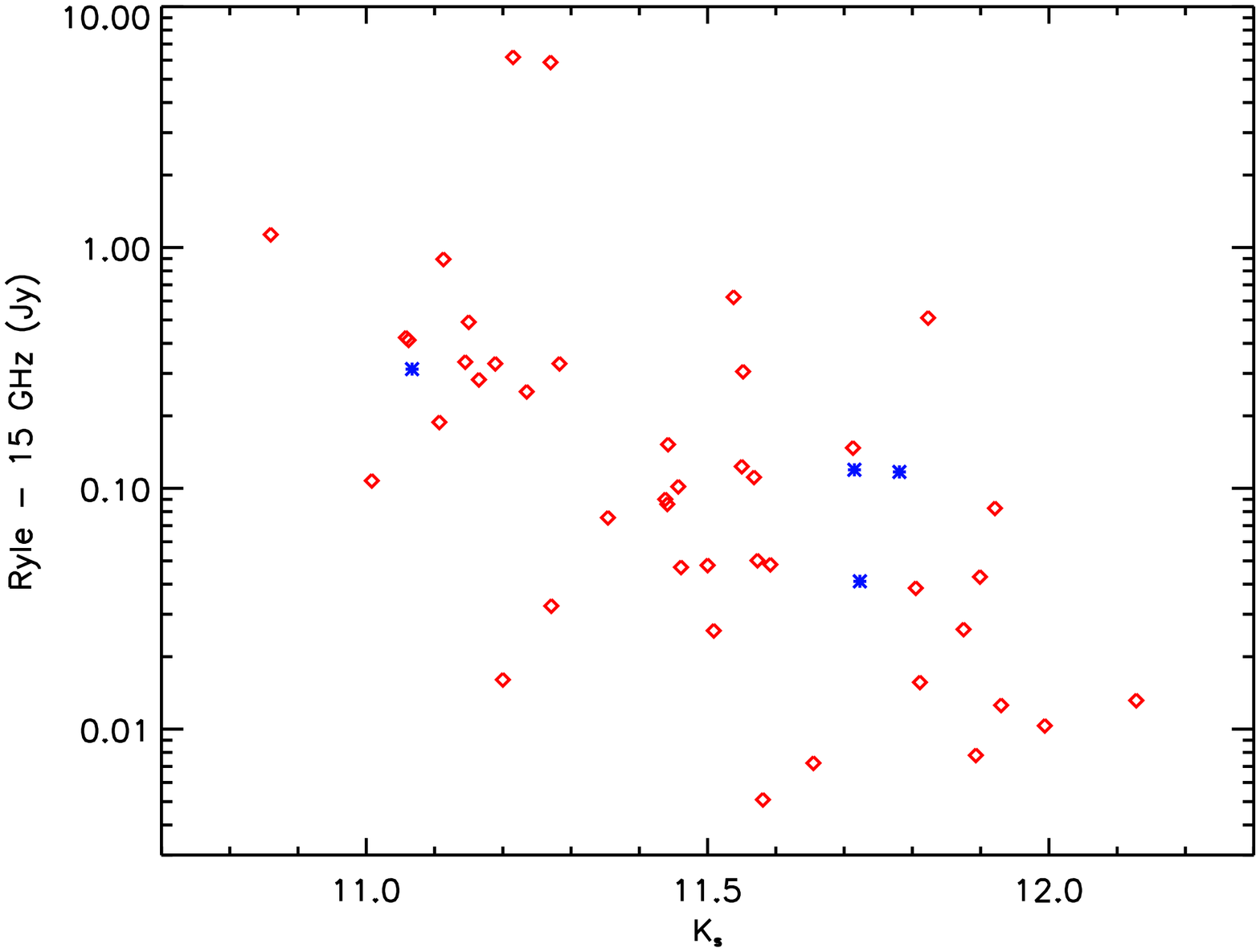}
\caption{{\it left:} Lightcurves of the infrared (red), SXR (green) and HXR 
(blue) during times of gamma-ray emission (dotted vertical lines) and a 1 Jy 
radio flare (dashed vertical line). {\it right:} Infrared magnitude ($\rm K_s$) 
vs. radio flux density (15 GHz). Note the logarithmic y-axis. The red points 
were taken during X-ray flaring states and the blue points during X-ray 
quiescence.}
\label{fig2}
\end{center}
\end{figure}

\subsection{Infrared Emission}

Because of the heavy absorption, due to interstellar and intrinsic material,
Cygnus X-3 is not observable in the optical and UV.  For this reason the
infrared is an important link between the radio and X-ray.

It was from infrared spectra that the mass-donating companion of Cygnus X-3 was 
identified to be a Wolf-Rayet star \cite{vk1}.  Infrared observations show the
4.8 hr orbital modulation similar to that seen in the X-ray, although it is not 
as strong (10\% as compared to 50\% in the X-ray) \cite{mcw}.  There have been
observations for which the infrared modulation has not been present \cite{bee}.
Additionally there have been observations of short intense (factor of 2) flares
imposed on top of the orbital modulation \cite{mcw}.

To improve our coverage in the infrared as part of our multi-wavelength campaign
we are observing in the near infrared using the PAIRITEL telescope.  PAIRITEL 
is the 1.3m automated telescope used for the 2MASS survey.  The final image 
mosaic created from an observation yields a 10 arcminute field with arcsecond 
resolution.  For Cygnus X-3 simultaneous ($\rm J,~H,~and~K_s$) observations are
achieved with a net exposure of $\rm \sim 600~s$.

{\tt Cygnus X-3 Infrared Spectral Changes:} An initial examination of the 
infrared shows that as Cygnus X-3 changes state (quiescent to flaring), the 
infrared spectrum changes.  An infrared color - color plot ($\rm H-K_S~vs.J-H$) 
is shown in Fig. 3 which shows a clear statistically significant correlation 
(see Table 2).  In Fig. 3 the average infrared spectrum is shown during 
quiescent (blue) and flaring (red) states.  While the flux in all bands rises in
the flaring state it is important to note the H band ($\rm \sim 1.6 \mu$)
experiences a noticeably larger increase.  This likely represents an increase in
one of the emission components that make up the infrared emission.

\begin{figure}
\includegraphics[angle= 90.0,width=0.5\textwidth]{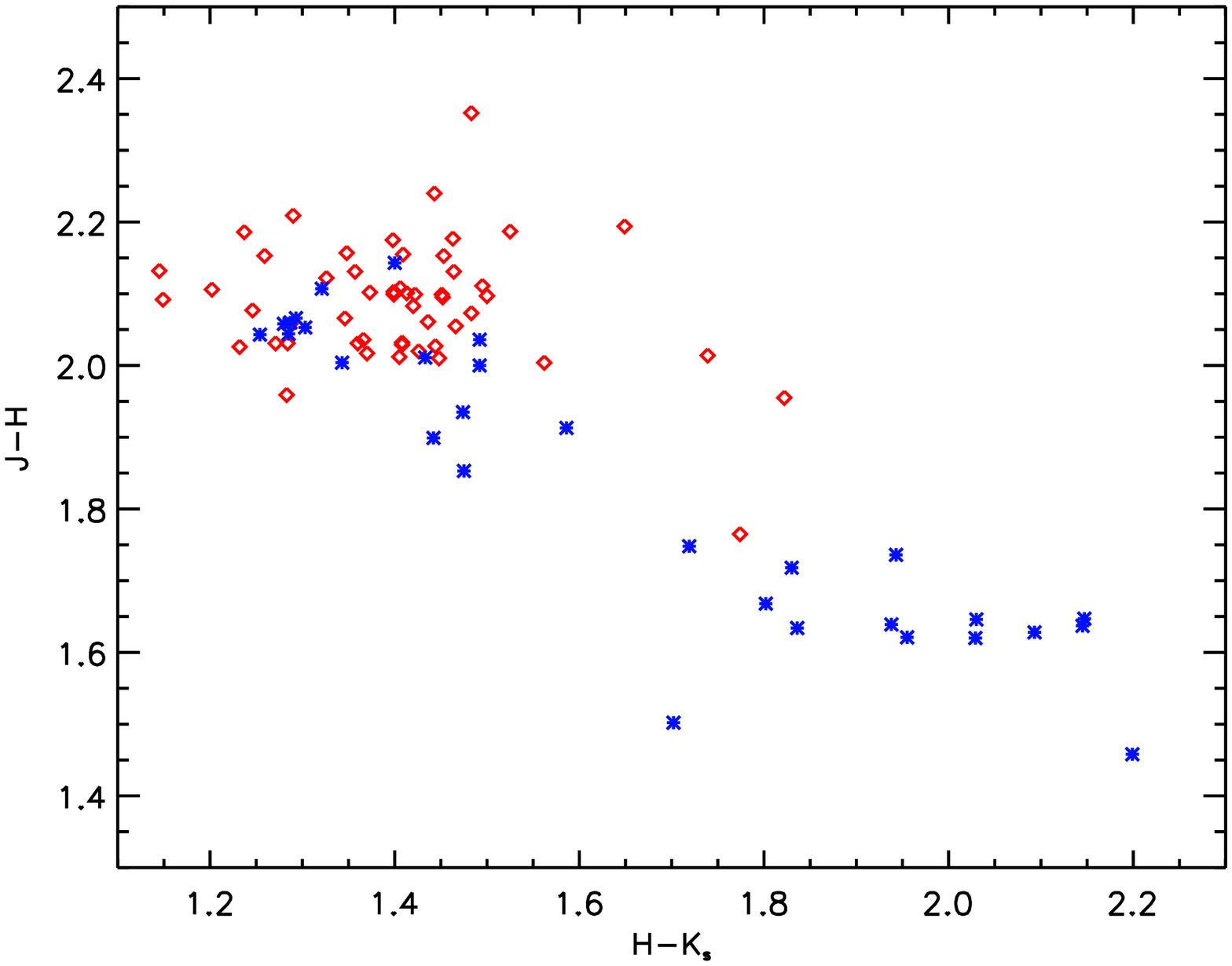}
\includegraphics[angle= 90.0,width=0.5\textwidth]{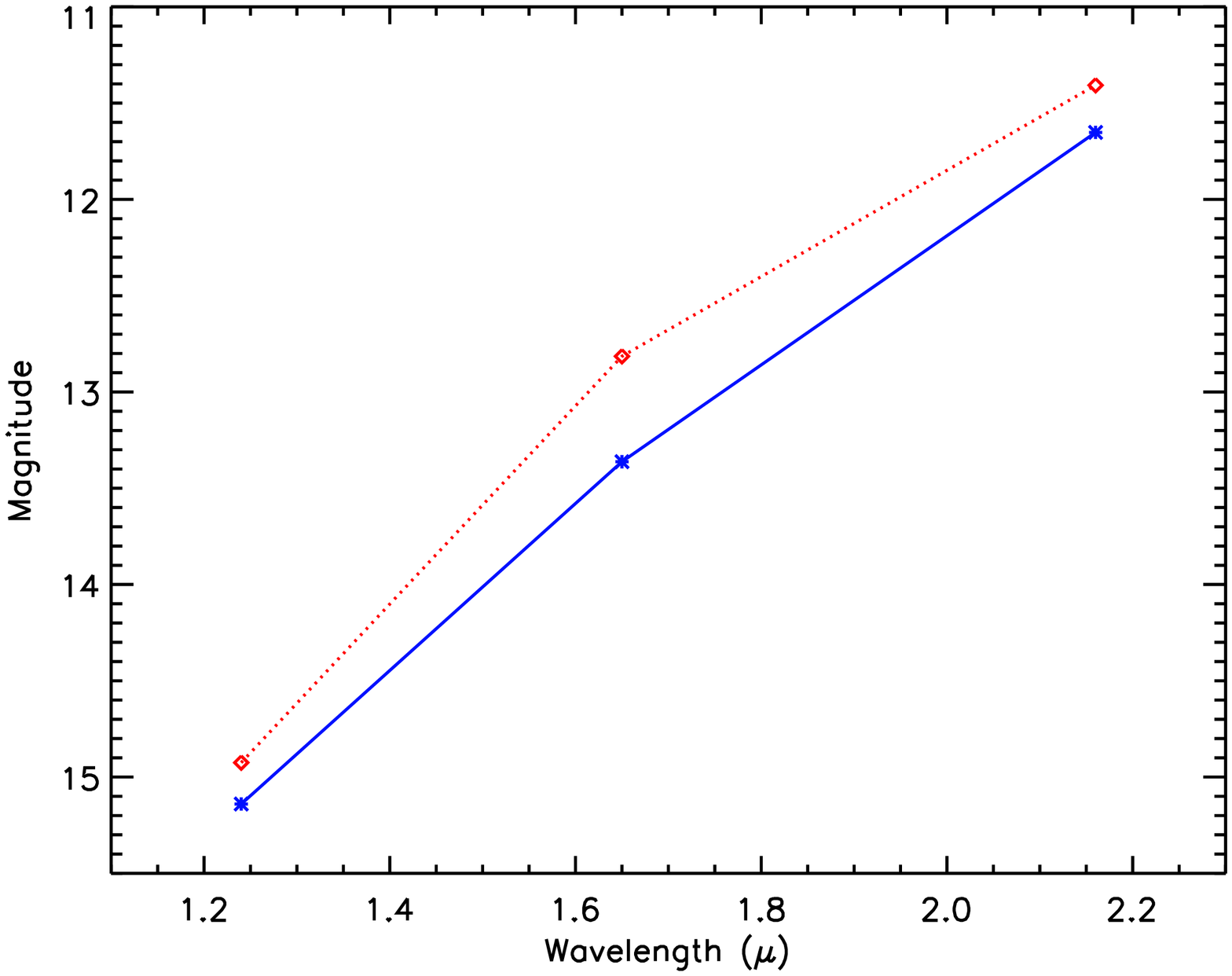}
\caption{{\it left:} An infrared color-color plot.  Note the correlation between
the colors which is heavily driven by the state changes.  The red represent
observations during a flaring state and blue during a quiescent state.
{\it right:} The average spectrum for the flaring (red) state and quiescent
(blue).  The wavelength is that of the band center.}
\label{fig3}
\end{figure}

{\tt SXR - HXR - Infrared Relationships:}  A comparison between the infrared and
the SXR/HXR show a clear correlation (SXR) and anti-correlation (HXR) with 
the infrared flux.  Fig. 4 shows both of these relations with the H band
(also see Table 2).  As the SXR increases the infrared flux increases and the
HXR flux decreases.  These correlations appear to be almost completely driven 
by the state change.

\begin{figure}
\begin{center}
\includegraphics[angle= 90.0,width=0.49\textwidth]{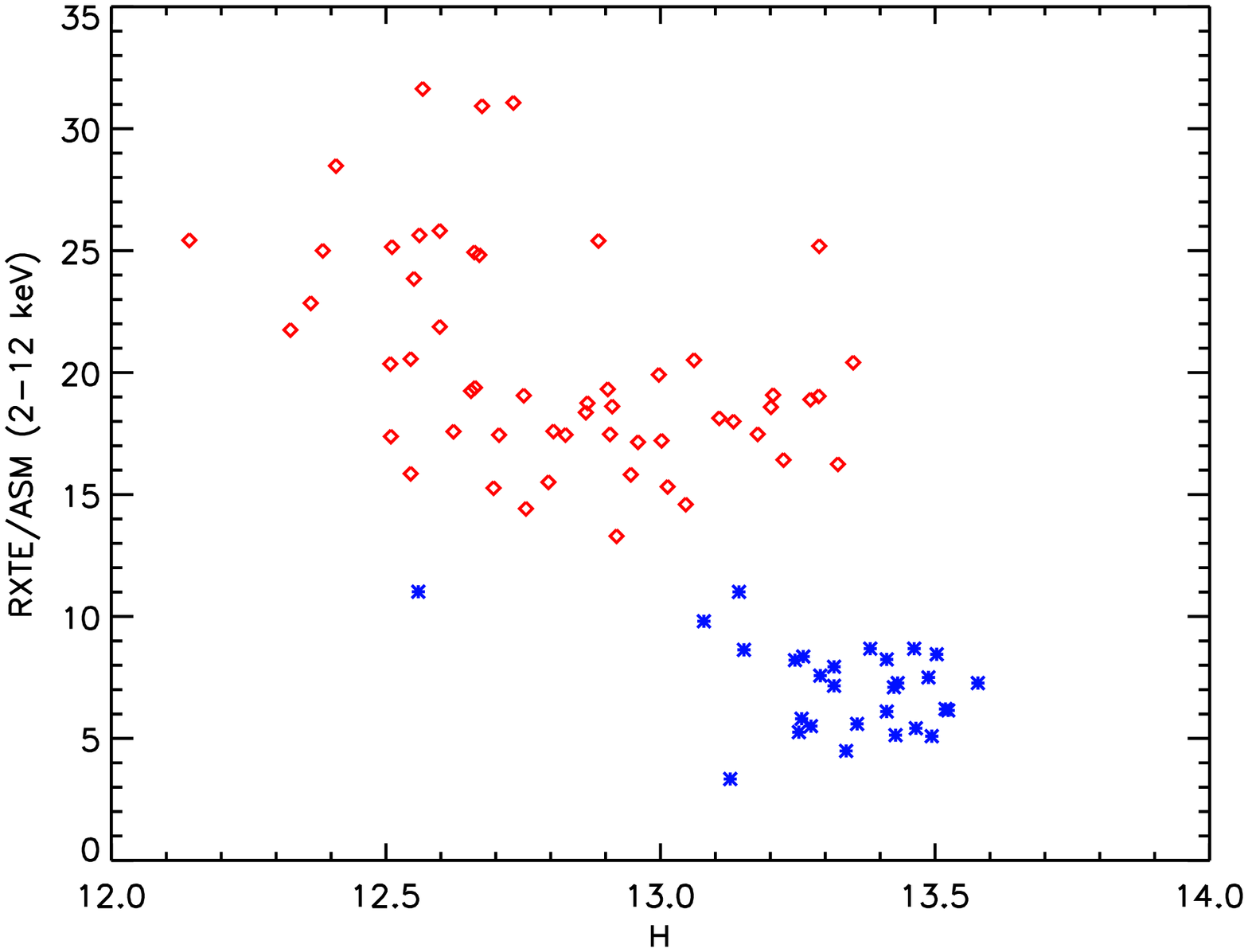}
\includegraphics[angle= 90.0,width=0.49\textwidth]{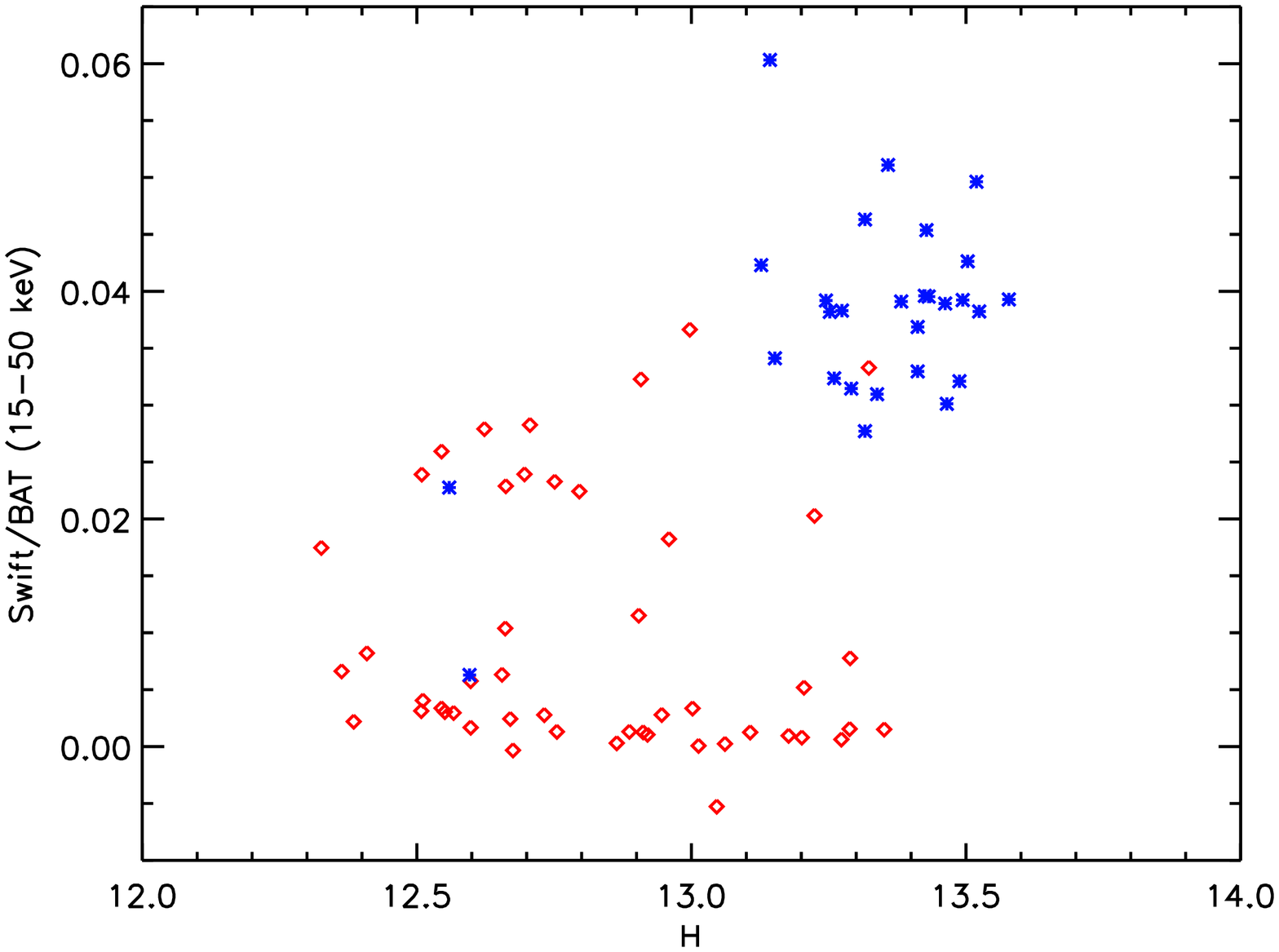}
\caption{{\it left:} Plot of the H band magnitude vs. SXR. Red are flaring state
observations and the blue are quiescent. {\it right:}  Plot of the H band 
magnitude vs. HXR. Red are flaring state
observations and the blue are quiescent.}
\label{fig4}
\end{center}
\end{figure}

{\tt Infrared - Radio Correlation:}  Infrared observations were also matched
with radio observations (maximum separation in time of $\rm <~day$) . Testing
relations between the various infrared bands and the radio show a correlation 
of the radio flux density 
with the $\rm K_s$ (see Fig. 2), a weak correlation with the H band, and a 
marginal correlation with the J band (see Table 2).  One implication of this is
that the synchrotron emission  responsible for the radio emission extends down 
into the infrared with its strongest contribution in the longer wavelength 
band.

{\tt Infrared - Gamma-Ray Relationship:}  During the 2010 period of gamma-ray
activity, infrared monitoring was also in place.  In Fig. 2 are 
the infrared observations, SXR daily observations, and the HXR semi-daily 
observations during this time period.  Labeled are periods of gamma-ray
emission and the time of a 1 Jy radio flare.  
Strikingly, the second period of gamma-ray emission, for which there is 
associated a radio flare, the infrared shows a direct response.
The infrared starts to increase in the 
$\rm \sim 14~days$ period prior to the radio flare.  Right after the peak of 
the radio flare the infrared shows a sudden decrease.  
There appear to be direct relationships between periods of gamma-ray 
emission, radio flares and infrared in Cygnus X-3.

\begin{table}
\begin{center}
\begin{tabular}{|c|c|c|c|}
\hline
Band/Color & Band/Color & $\rm r_s$ & $\sigma$ \\
\hline
 $\rm H-K_s$ & $\rm J-H$ & -0.519 & $\rm 3 \times 10^{-7}$ \\
\hline
 $\rm H$ & $\rm SXR$ & -0.738 & $\rm 1 \times 10^{-15}$ \\
\hline
 $\rm H$ & $\rm HXR$ & 0.542 & $\rm 4 \times 10^{-7}$ \\
\hline
 $\rm K_s$ & $\rm Radio  $ & -0.640 & $\rm 1 \times 10^{-5}$ \\
\hline
 $\rm H$ & $\rm Radio  $ & -0.558 & $\rm 3 \times 10^{-4}$ \\
\hline
 $\rm J$ & $\rm Radio  $ & -0.513 & $\rm 1 \times 10^{-3}$ \\
\hline
\end{tabular}
\caption{Infrared correlations: For different bands/colors Spearman rank-order
correlation test was preformed.  For each case the $\rm r_s$ is the Spearman 
rank coefficient and $\sigma$ is significance (probability of no correlation)
are given.}
\label{tab2}
\end{center}
\end{table}

\section{Summary}

Through multi-wavelength campaigns of Cygnus X-3 important insights 
are being gained.  Cygnus X-3 has been shown to be one of the few XRBs to produce 
gamma-ray emission.  This emission is coupled to state behavior and major radio 
flares.  The infrared is providing an important bridge between the radio and the
X-ray. These studies are allowing us to probe some of the most extreme 
environments in nature: high mass flow in a strong gravitational field.

\end{document}